\documentclass{article}

\usepackage{arxiv}

\usepackage[utf8]{inputenc}
\usepackage[T1]{fontenc}
\usepackage{url}
\usepackage{multirow}
\usepackage{booktabs}
\usepackage{amsfonts}
\usepackage{nicefrac}
\usepackage{microtype}
\usepackage{lipsum}
\usepackage{amsmath}
\usepackage{graphicx}
\usepackage[symbol]{footmisc}
\usepackage{caption} 
\graphicspath{ {images/} }

\title{Graph convolutional neural networks as "general-purpose" property predictors: the universality and limits of applicability}

\begin{document}

\begin{center}
 {\LARGE Graph convolutional neural networks as "general-purpose" property predictors: the universality and limits of applicability}
\end{center}

\begin{center}
 Vadim Korolev\footnote[1]{\textit{Email address}: \texttt{korolev@colloid.chem.msu.ru}}\textsuperscript{1,2}, Artem Mitrofanov\textsuperscript{1,2}, Alexandru Korotcov\textsuperscript{1}, Valery Tkachenko\textsuperscript{1}
\end{center}

\begin{center}
 \textsuperscript{1}\textit{Science Data Software, LLC, 14909 Forest Landing Cir, Rockville, MD 20850, USA}\\
 \textsuperscript{2}\textit{Lomonosov Moscow State University, Department of Chemistry, Leninskie gory, 1 bld. 3, Moscow 119991, Russia}
\end{center}

\vspace{3mm}

\begin{abstract}
Nowadays the development of new functional materials/chemical compounds using machine learning (ML) techniques is a hot topic and includes several crucial steps, one of which is the choice of chemical structure representation. Classical approach of rigorous feature engineering in ML typically improves the performance of the predictive model, but at the same time, it narrows down the scope of applicability and decreases the physical interpretability of predicted results. In this study, we present graph convolutional neural networks (GCNN) as an architecture that allows to successfully predict the properties of compounds from diverse domains of chemical space, using a minimal set of meaningful descriptors. The applicability of GCNN models has been demonstrated by a wide range of chemical domain-specific properties. Their performance is comparable to state-of-the-art techniques; however, this architecture exempts from the need to carry out precise feature engineering.
\end{abstract}


\section{Introduction}
The design of new functional materials is associated with significant computational costs. The use of different approximations substantially based on density functional theory\cite{Hohenberg1964,Kohn1965} (DFT) makes it possible to solve Schrödinger equation numerically without critical loss of accuracy for calculation of ground-state properties. However, DFT calculations are still disappointingly time-consuming for materials and molecular design, especially if we consider their extensive use for large databases\cite{Curtarolo2012a,Saal2013,Jain2013}. An alternative approach combines the quantitative structure-property relationship (QSPR) modeling and ML techniques for materials property predictions based on their structure. This method accelerates the calculation of properties by several orders of magnitude. Nevertheless, it remains unclear how widely this approach can be applied and replace the DFT calculations.

It is well known that appropriate representation of considered structures plays a crucial role in property prediction using machine learning\cite{Mitchell2014,Ghiringhelli2015,Seko2017,Ramprasad2017}. Design of input data, and so called \textit{feature engineering}, represents a challenge. In particular, a specific feature vector has to be constructed for each use case; the more specific the target property, the more specific the feature vector should be. All these issues prevent the use of one unified approach for different domains of chemical space.

Nevertheless, several attempts have been made to create universal models that do not require precise feature engineering. The Gaussian process regression\cite{Csanyi2017} and deep tensor neural networks\cite{Schutt2018} were used as basic algorithms. However, it should be noted that the priority attention in these studies was paid to the prediction of the thermodynamic stability of molecules/materials, i.e., their energy. State-of-the-art results have been achieved using graph-based neural networks\cite{Chen2019} on large datasets generated in a high-throughput manner (e.g., QM9 for molecules, Materials Project database for inorganic crystals). At the same time, many physicochemical quantities cannot be obtained as a result of DFT calculations. Data scarcity makes it very difficult to build highly efficient predictive models (including those based on graph-based neural networks) in the case of small experimentally obtained datasets. Another issue of particular interest is a generalization of such universal models to porous or low-dimensional materials, which represent an intermediate case between molecular and crystalline systems from an atomic connectivity perspective. However, the above difficulties have not been considered in detail.

This study is devoted to the development of a universal approach that would allow predicting the properties of both molecules and materials with varied atomic connectivity, using simple, physically interpretable descriptors. A graph-based convolutional neural network is considered here as a base architecture for universal property prediction. Similar architectures were successfully used to predict the properties of structures from distinct domains of chemical space\cite{Schutt2017,Faber2017,Xie2018}. To the best of our knowledge, this work also represents the first attempt to apply graph-based NNs for the prediction of two-dimensional and porous materials’ properties. The limits of applicability of graph-based NNs, as a flip side of their universality, are also clearly demonstrated.

\section{Methods}
Convolutional neural networks (CNNs) are one of the most promising and fast-developing classes of machine learning algorithms due to their exceptional performance in image, video and speech recognition tasks\cite{Lecun2015}. Ordinary CNNs (strictly speaking, convolutional kernels) require input data to be on the regular grid. Unfortunately, most of the real use cases deal with highly unordered data, which can be sometimes represented as a graph. Extension of convolutional kernels to an irregular domain is a nontrivial task, thus in recent years the efforts of several scientific groups have focused on its solution\cite{Defferrard2016,Duvenaud2015}. A graph-based architecture of artificial neural networks is known as a graph convolutional neural network (GCNN).

The attractive idea is to adapt and apply graph neural networks for materials property prediction since chemical structures can be represented in the form of graphs with atoms for nodes and bonds for edges. Previously, this concept was used mainly for organic molecules, and so far only a few studies have been devoted to their application for periodic structures (inorganic crystals) property prediction\cite{Chen2019,Xie2018}. For the sake of universality, we propose to use only the adjacency matrices and the set of most common properties of corresponding element/lattice site as input data for the GCNN model. More detailed information on GCNN architecture is provided in Supporting Information.

\begin{figure}[h]
  \centering
  \includegraphics[height=7cm]{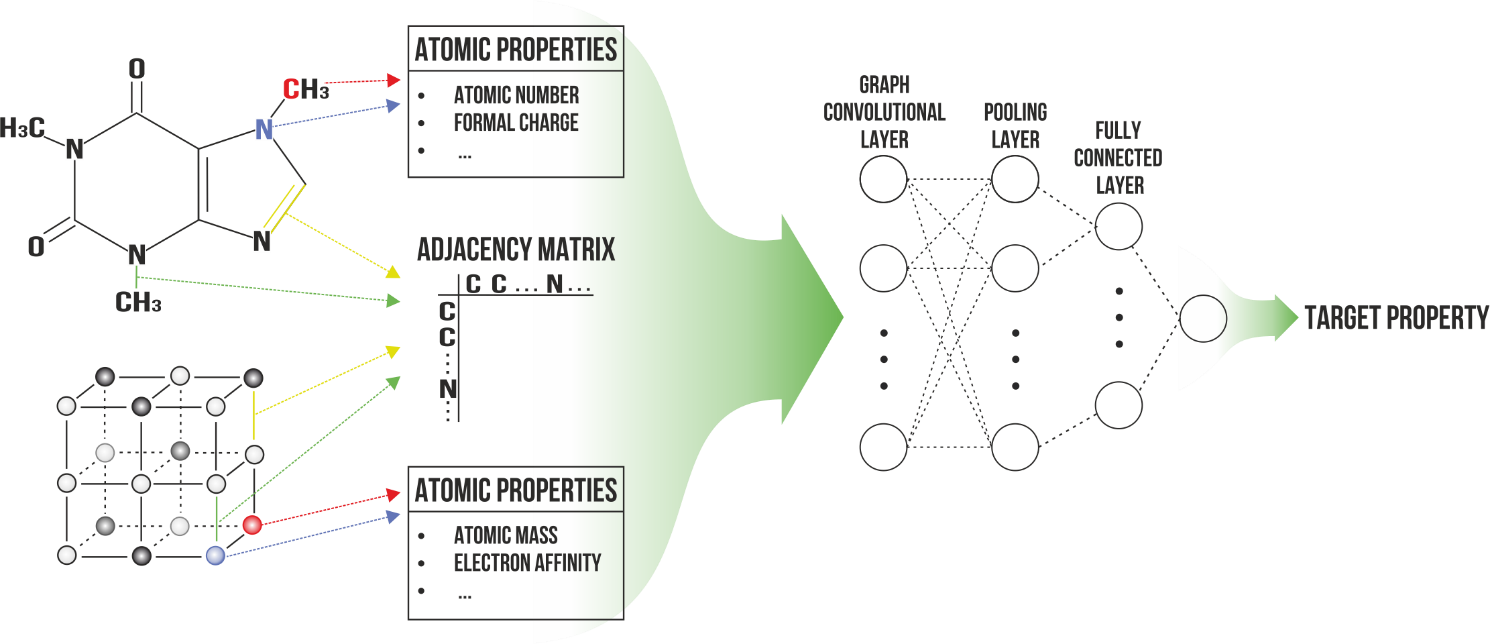}
  \caption{General workflow of training process.}
  \label{fig:fig_1}
\end{figure}

Two slightly different sets of descriptors for non-periodic and periodic structures were used to represent structures, and each of them reflects the specificity of the above systems. For non-periodic structures (organic molecules), only node-wise descriptors were used, and structural features were excluded entirely from consideration, which is opposite to the case of periodic structures where those features were included (for more details see Supporting Information).

It should be noted that we purposely excluded from consideration linear combinations of the mentioned descriptors. Such artificially generated quantities are widely used in conjunction with random forest/gradient boosting models, with the subsequent selection of the most valuable features. In fact, this methodology is one of the variations of feature engineering. However, as stated earlier, our goal was to build the simplest, most universal model, without the need to use extensive sets of domain-specific features and their combinations.

To demonstrate the universality and limits of applicability of GCNN models we train models on multiple diverse datasets, which are covering wide range of various parameters:

\begin{itemize}
  \item Dimensionality – 0D (organic molecules), 2D (layered materials), 3D (all others).
  \item Structural diversity – high (porous materials), medium/low (all others).
  \item Compositional diversity – low (porous materials), medium/high (all others).
\end{itemize}

\textbf{Molecules.} Previously, similar graph-based neural networks demonstrated high accuracy of prediction of multiple molecular properties related to different levels, including quantum mechanical, physical chemistry, biophysical and macroscopic physiological properties\cite{Wu2018}. Only a few datasets from the last three levels were used to confirm the performance of GCNN models.

Water solubility models were trained using a dataset\cite{Huuskonen2000} containing 1299 molecules with logS = – 5 taken as a cut-off value for classification. We also trained a number of models for biological endpoints, including human ether-a-go-go-related gene (hERG) inhibitors dataset\cite{Wang2012} with 373 active and 433 inactive molecules; 175 active, 19 604 inactive molecules from CDD Public datasets\cite{Guiguemde2010,Gamo2010,GagaringK.;BorboaR.;FrancekC.;ChenZ.;Buenviaje} (malaria); and dataset with 253 active, 69 inactive molecules that have been scored for probe-likeness by medical chemists\cite{Litterman2014}.

\textbf{Inorganic crystals.} Experimental determination of inorganic crystal properties is a complex task due to difficulties with a monocrystalline synthesis. In this study, we used datasets with DFT calculated properties to train models. Two sources of data were used: AFLOWlib\cite{Curtarolo2012b} and JARVIS-DFT\cite{Choudhary2018a} databases. To train and validate models for prediction of thermomechanical properties, datasets with 2748 and 770 compounds, respectively, were taken from Isayev et al\cite{Isayev2017}. Six properties include the bulk modulus, shear modulus, Debye temperature, heat capacity at constant pressure, heat capacity at constant volume, and the thermal expansion coefficient were calculated with the AEL–AGL integrated framework\cite{Toher2014}. Metal/insulator and magnetic/non-magnetic classification tasks were performed on 25468 and 25131 compounds, extracted from AFLOWlib and JARVIS-DFT databases, respectively. Optimized two-dimensional structures and corresponding exfoliation energies for 601 compounds were extracted from JARVIS-DFT\cite{Choudhary2017} database. All calculations were provided with PBE and optB88 functionals.

\textbf{Porous crystalline materials.} To demonstrate the applicability of GCNN models for porous materials’ property prediction we consider two domain-specific properties: the bulk and shear modulus of pure-silica zeolites from IZA database and Xe/Kr infinite dilution selectivity of Computation-Ready, Experimental (CoRE) Metal-Organic Frameworks (MOF).

To predict the bulk and shear moduli of pure silica zeolites we use a subset of the Database of Zeolite Structures presented by Coudert\cite{Coudert2013}. Seven of 122 zeolite frameworks were excluded from consideration since errors occurred during the generation of the corresponding structural descriptors (see Supporting information). B3LYP hybrid exchange-correlation functional was used to obtain the elastic data, isotropic values of bulk and shear moduli were calculated using Voigt-Ruess-Hill averages.

Xe/Kr adsorption data for all MOF structures (for which density derived electrostatic and chemical charges have been obtained) was modeled via a classical force field (FF), namely, a universal force field\cite{Witman2017}. To demonstrate the influence of intrinsic flexibility on the Henry regime adsorption properties in CoRE MOF structures all calculations were carried out both in a rigid and in a flexible approximation.

\section{Results and discussion}

\textbf{Molecules.} Like many other breakthroughs in chemoinformatics, the deep learning revolution in the field was mainly caused by the needs of drug design\cite{Gawehn2016,Mamoshina2016,Goh2017,Aspuru-Guzik2018}. Graph-related neural networks are no exception.  To date, they are mainly used to predict molecular properties\cite{Faber2017,Kearnes2016}. Recently, several studies present the large-scale comparison of various ML techniques, including graph-based neural networks, proving their high performance\cite{Wu2018,Mayr2018}. Therefore, we take into consideration only a few molecular datasets to prove the concept.

We compare the prediction performance of GCNN models against the prediction performance of the support vector machine (SVM) and the feed-forward neural networks (FNN) models on four molecular datasets with ECFP6\cite{Rogers2010} fingerprints. Following the original methodology\cite{Korotcov2017}, we test our models on an external test set (20\% compounds from initial set) with five-fold cross validation. The results are presented in Table \ref{table:1}. These two algorithms have been chosen as the most efficient – 5-layer FNN and SVM (among all classic machine learning algorithms) rank above all other presented ML methods\cite{Korotcov2017}.

\begin{table}[h!]
\caption{Summary of performances (molecules-related tasks): GCNN models versus SVM and FNN models.}
\label{table:1}
\centering
\begin{tabular}{ ||c||c|c|c|c|c|c| }
\hline
\multirow{ 2}{*}{} & \multicolumn{2}{|c|}{GCNN (this study)} & \multicolumn{2}{|c|}{SVM\cite{Korotcov2017}} & \multicolumn{2}{|c|}{FNN\cite{Korotcov2017}} \\
\cline{2-7}
& ROC AUC & Accuracy & ROC AUC & Accuracy & ROC AUC & Accuracy \\
\hline
solubility & 0.97 & 0.96 & 0.93 & 0.90 & 0.93 & 0.93 \\
hERG & 0.85 & 0.78 & 0.86 & 0.80 & 0.84 & 0.80 \\
malaria & 0.93 & 0.99 & 0.90 & 0.97 & 0.97 & 0.99 \\
probe-like & 0.62 & 0.76 & 0.66 & 0.76 & 0.56 & 0.77 \\
\hline
\end{tabular}
\end{table}

The most obvious tendency deals with data scarcity and class imbalance. GCNN models outperform or perform similarly as other algorithms on well-balanced datasets (solubility and hERG). Nevertheless, GCNN models as a particular case of graph-based models are not robust enough\cite{Wu2018,Mayr2018} to perform well on highly imbalanced (malaria, active/inactive ratio is 0.0089) and small datasets (322 samples in total).

As an intermediate conclusion, we may denote GCNN as a convenient tool for the purposes of molecule QSPR modelling. The accuracy of the approach is comparable with well-known and reliable methods, though it can’t be called a breakthrough in regards to small organic molecules.

\textbf{Inorganic crystals.} AFLOWlib\cite{Curtarolo2012b} and JARVIS-DFT\cite{Choudhary2018b} databases have already been used to develop ML models for prediction of thermomechanical properties. In contrast to this study, the available machine-learning frameworks are based on precise feature engineering and algorithms with high interpretability, such as gradient boosting decision trees (GBDT). Following the original methodology\cite{Isayev2017}, we test our models on an external test set ($\sim$20\% compounds from initial set) with a five-fold cross validation. Table \ref{table:2} contains the performance metrics for six regression models that predict thermomechanical properties of bulk materials and exfoliation energy of potentially exfoliable 2D-layered materials. In most cases, GBDT models slightly outperform GCNN models. Furthermore, we also develop two predictive models for metal/insulator and magnetic/non-magnetic classification tasks. The area under the ROC curve and accuracy for the two classification tasks obtained with GCNN/GBDT models were used to evaluate and compare models. These models show similar accuracy with the area under the curve at 0.97/0.98\cite{Isayev2017} for metal/insulator and 0.94/0.96\cite{Choudhary2018b} for magnetic/non-magnetic classification tasks respectively.

Surprisingly, that GCNN models demonstrate sufficient accuracy not only for bulk materials property prediction but also for initial screening of potentially exfoliable materials. Following the original methodology\cite{Choudhary2018b}, we test our models on an external test set (10\% compounds from initial set) with five-fold cross validation. Our GCCN model has MAE for exfoliation energy (37 meV/atom) that comparable with a MAE of GBDT model, is significantly less than the threshold value for potentially exfoliable 2D-layered materials (200 meV/atom). Strictly speaking, only a few 2D materials are true monolayers. Most of them contain several layers in the direction perpendicular to the free surface, thus, the corresponding chemical graph used to predict exfoliation energy should be weighted, since the edges have an unequal contribution to the surface/exfoliation energy. Nevertheless, it should be concluded that elemental descriptors are sufficient for prediction of exfoliation energy with appropriate accuracy (3D Voronoi descriptors were excluded from consideration).

Due to the high interpretability of GBDT models, it is possible to rank the features of importance for the predictive model. According to Isavev et al \cite{Isayev2017}, the most important features are combinations of element properties, while within the framework of our GGNNs model, linear combinations of properties are not taken into account for the sake of model simplicity. Additionally , we do not use some of the specific properties also included in the descriptor list (effective atomic charge, chemical hardness, van der Waals radius, second and third ionization potentials). Moreover, structural representation used by Choudhary\cite{Choudhary2018b} contains very specific features – charge-based and classical force-field inspired descriptors, bond-angle and dihedral-angle distributions, etc. Thus, with a slight decrease of model performance, the total length of feature vector decreases from 2494 and 1557 for Isayev’s and Choudhary’s GBDT models, respectively, to 41 descriptors for our GCNN model.

\begin{table}[h!]
\caption{Summary of performances (materials-related tasks): GCNN models versus GBRT models.}
\label{table:2}
\centering
\resizebox{\textwidth}{!}{%
\begin{tabular}{ ||c||c|c|c|c|c|c| }
\hline
\multirow{ 2}{*}{} & \multicolumn{3}{|c|}{GCNN (this study)} & \multicolumn{3}{|c|}{GBDT\cite{Isayev2017}} \\
\cline{2-7}
& MAE & RMSE & R$^{2}$ & MAE & RMSE & R$^{2}$ \\
\hline
Bulk modulus, GPa & 15.21 & 24.61 & 0.91 & 12.00 & 21.13 & 0.93 \\
Shear modulus, GPa & 14.88 & 21.01 & 0.87 & 13.31 & 18.94 & 0.90 \\
Debye temperature, K & 50.97 & 71.21 & 0.91 & 42.92 & 64.04 & 0.93 \\
Heat capacity at constant pressure, $k_{b}$/atom & 0.07 & 0.10 & 0.91 & 0.06 & 0.10 & 0.92 \\
Heat capacity at constant volume, $k_{b}$/atom & 0.05 & 0.08 & 0.93 & 0.05 & 0.07 & 0.95 \\
Thermal expansion coefficient, K$^{-1}$ & 9.63$\times10^{-6}$ & 1,56$\times10^{-5}$ & 0.81 & 5.77$\times10^{-6}$ & 1.95$\times10^{-5}$ & 0.76 \\
Exfoliation energy, meV/atom & 36.5 & 69.4 & 0.22 & 37.3 & – & – \\
\hline
\end{tabular}
}
\end{table}

\textbf{Porous crystalline materials.} Recently, the bulk and shear moduli of pure-silica zeolites were calculated using five classical interatomic potentials\cite{Siddorn2015} (BKS\cite{VanBeest1990}, Catlow\cite{Sanders1984}, Gale\cite{Gale1998}, Matsui\cite{Tsuneyuki1988}, Sastre\cite{Sastre2003}). These results can serve as a ground-state level to probe the accuracy of other approaches (using machine learning, in our case), due to the wide applicability of classical model potentials for calculation of mechanical properties\cite{Combariza2013}.

Table \ref{table:3} contains performance metrics for force field models as mentioned above, based on five state-of-the-art interatomic potentials. Besides, the gradient boosting regressor (GBR) model was also used to predict the mechanical properties of zeolites from the same dataset\cite{Evans2017}. Following the original methodology\cite{Evans2017}, we test our models using three-fold internal cross validation. The values of the bulk $K$ and shear $G$ moduli obtained with GBR and GCNN models are also provided in Table 3.

The GCNN model significantly outperforms all force field models, but the accuracy of the GBR model is slightly better. According to Evans et al\cite{Evans2017}, the local descriptors (in particular, Si–O–Si angles and parameters related to the Si–O bonds) are the most crucial features for the prediction of mechanical properties. Due to the peculiarities of graph-based structure representation, it is not a trivial task to implement the descriptors associated with the statistical distribution for bonds and angles between atoms as node-wise features. Also, as an alternative approach, the secondary building blocks can be used as the vertices of the chemical graph instead of atoms. This insight is to be addressed in coming studies.

\begin{table}[h!]
\caption{Summary of performances (porous materials-related tasks): GCNN models versus GBR model and conventional model potentials.}
\label{table:3}
\centering
\begin{tabular}{ ||c||c|c| }
\hline
 & RMSE, $K$ & RMSE, $G$ \\
\hline
GCNN & 13.14 & 6.40 \\
GBR\cite{Evans2017} & 10.00 & 4.74 \\
BKS\cite{Siddorn2015} & 22.7 & 36.1 \\
Catlow\cite{Siddorn2015} & 18.8 & 11.7 \\
Gale\cite{Siddorn2015} & 20.0 & 12.6 \\
Matsui\cite{Siddorn2015} & 16.8 & 29.4 \\
Sastre\cite{Siddorn2015} & 18.1 & 14.1 \\
\hline
 & RMSE, $S_{f}$ & RMSE, $S_{r}$ \\
 \hline
GCNN & 3.98 & 5.32 \\
\hline
\end{tabular}
\end{table}

Furthermore, Database of Zeolite Structures is a typical example of small materials dataset, and, in a sense, it is similar to the previously discussed probe-like dataset. Data scarcity imposes principal limits on the level of accuracy of ML models regardless of structure representation. Only advanced techniques, such as incorporating the crude estimation of property (CEP) in the feature vector\cite{Zhang2018}, can significantly improve the performance of implemented models.

Table \ref{table:3} also contains performance metrics of GCNN models predicted infinite dilution selectivity of Xe/Kr in rigid $S_{r}$ and flexible $S_{f}$ approximations. As in the previous case, GCNN models have low accuracy in prediction of porous materials’ properties as opposed to other subdomains of chemical space. The models for prediction of infinite dilution selectivity in flexible mode show even better performance, but due to the low accuracy of both models the result is not significant. Nevertheless, considering the ratio between the range of selectivity values and corresponding RMSE, GCNN models enable at least a meaningful qualitative comparison of most promising candidates for adsorption-related applications. Furthermore, graph-based representation is suitable for modeling of strong interatomic interactions, at the same time it is well known that van der Waals interactions play a significant role in MOFs and similar classes of porous materials. Also, the interaction of atoms only with the first coordination sphere were taken into account for GCNN (see Supporting information). This restriction reduces the time required for training the model but also makes it challenging to consider long-range forces. This approach can be compared to using a short cutoff radius in the Bayesian GPR framework\cite{Csanyi2017}.

In contrast to the properties of crystalline materials, a network of pores primarily determines the properties of porous materials associated with adsorption. As has been shown, even only-structural descriptors that ignore entirely chemical diversity are suitable for the clustering of porous materials\cite{Lee2017}. Given the specifics of these properties, pore-centered descriptors, instead of atom-centered ones, used successfully to predict the properties of crystalline materials, seem to be a more appropriate choice.

\section{Conclusions}

In this study, we have shown that GCNN architecture with a minimal set of interpretable descriptors could be a universal tool for fast initial screening and search of perspective materials from the various subdomains of chemical space. Its performance was tested with a broad set of chemical domain-specific properties, including biological activity for organic molecules, thermomechanical properties for inorganic crystals, exfoliation energy for potentially exfoliable materials, elastic moduli and infinite dilution selectivity of Xe/Kr for porous materials. Except for the last subdomain, the GCNN models demonstrate excellent accuracy, which is comparable with the best known approaches. However, even for the domain of porous materials GCNN models can be applied for an initial search of advanced materials for specific applications. GCNN models are still vulnerable to the effects of data scarcity and class imbalance data, but at the same time, graph-based NNs demonstrate state-of-the-art performance on well-balanced datasets with a sufficient number of samples.

\section{Acknowledgments}

This work was supported by the Russian Science Foundation (project no. 19-73-20115).

\end{document}